\documentstyle[epsfig,12pt]{article}
\def\kpipi{$D^+\rightarrow K^-\pi^+\pi^+$\ }

\def\ka14{$K^*_0(1430)$\ }
\def\d3pi{$D^+\rightarrow \pi^-\pi^+\pi^+$\ }
\def\ds3pi{$D_s^+\rightarrow \pi^-\pi^+\pi^+$\ }

\begin{document}

\title{ Complex amplitude phase motion  
in Dalitz plot heavy  meson three body decay. }

\author{Ignacio Bediaga and Jussara M. de Miranda \\ Centro 
Brasileiro de Pesquisas F\'\i sicas, \\ Rua Xavier Sigaud 150, 22290-180 
-- Rio de Janeiro, RJ, Brazil\\ bediaga@cbpf.br and jussara@cbpf.br}

\maketitle

\begin{abstract}
We propose a  method to determine the phase motion of
a complex amplitude in three body heavy meson decays. We show 
that the phase variation of a complex amplitude can be directly revealed through
the interference observed in the Dalitz Plot region where it crosses with a
well established resonant state. This method  could be 
applied to the decays \d3pi and $D^+\rightarrow K^-\pi^+\pi^+$, 
to determine  whether  the low mass states $\sigma$  and $\kappa$, suggested by
E791, have phase motions compatible with  resonances.

\end{abstract}

\vspace{3cm} 
\begin{tabbing}

\=xxxxxxxxxxxxxxxxxx\= \kill

\>{\bf Keywords:} \> Heavy Meson Decay; Dalitz plot;Scalar Mesons; Resonances. \\

\>{\bf PACS Numbers:} \> 13.25.Ft, 14.40.Ev, 14.40.Cs,11.80.Et.

\end{tabbing}

\newpage
Recently the Fermilab experiment E791 with Dalitz plot analysis showed 
strong evidence for 
the existence of two light and broad scalar resonances. One resonance is
compatible with the  isoscalar meson $\sigma$, and is observed  
in Cabbibo suppressed decay \d3pi  \cite{e791_1}.  The  
 other, its  strange  partner, the meson $\kappa$ seen in the 
 \kpipi decay \cite{e791_2}. These results  opened  a  
 new experimental window for understanding scalar meson 
 spectroscopy \cite{torn}.
 
 In these analyses, each possible resonance  amplitude is  
 represented by Breit-Wigner functions multiplied by angular distributions
 associated with the spin of the resonance. 
 The various contributions are combined in a coherent sum with complex
 coefficients that are  extracted from maximum likelihood fits to the data. The absolute value of the
 coefficients are related to the relative fraction of each contribution and the
 phases takes into account final state interaction (FSI) between the resonance and the
 bachelor pion
  \footnote{ This phase is considered constant because it depends only on
  the total energy of the system, i.e. the heavy meson mass.}.
 To fit E791 data, it was necessary to  include extra, not previously observed
 scalar resonances. For the new states, modeled by Breit Wigner amplitudes,
  they measure mass and width. The \d3pi
 decay required a scalar with mass 
  $478^{+24}_{-23} \pm 17 $ MeV/$c^2$ and width 
 $324^{+42}_{-40} \pm 21 $ MeV/$c^2$.
 The high statistics sample of \kpipi needed a scalar with  
 mass of $797\pm 19\pm 43$~MeV/c$^2$ and width 
$410\pm 43\pm 87$~MeV/c$^2$  for a good confidence level fit. In
both cases, the extra contribution is dominant, accounting for  approximately
 half of the decay.
Due the importance of these mesons in many 
areas of particle and nuclear physics, it is desirable 
to be able to have  confirmation with 
a direct observation of their phase motion, without having to assume a priori 
the Breit-Wigner phase variation \cite{ochs}.  

Partial wave analyses (PWA)     have been  used  successfully     
in  hadron-hadron   scattering 
  to observe the phase motion of complex amplitudes.  For example, the LASS 
  experiment \cite{lass} did a partial wave analysis of  the  $K^- \pi^+$ 
  spectrum  in   $ K^- p \to K^- \pi^+ n  $ interaction 
  with transverse momentum less than $\ 0.2$ (GeV/c)$^2 $. The low momentum transfer assures
 first that one $\pi^+$ meson exchange 
between the incoming $K^-$ and the proton is dominant, and second that
 there is no important contribution to final state interactions 
between the  $K^- \pi^+$ system and the outgoing neutron. 
This is a physical situation completely different from charm three body 
decay, where all phase space is taken into account,
including low and high momentum transfer 
between  the resonance state  and the bachelor particle.  The 
strength of  the resonance-bachelor particle interaction 
  in charm decay is given by the FSI  phase. In fact, 
 experimental results show a tendency of large values for the strong  phase, thus
non negligible FSI \cite{e791_1,e791_2,e791_3,cleo, e687-1,e687-2}.

\begin{figure}[hbt]
\epsfxsize=18pc
\centerline{
\epsfbox{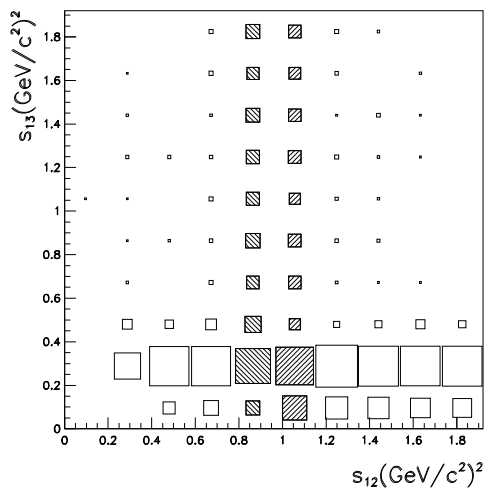}}
\caption{Dalitz plot distribution for two crossing scalar mesons,  
$f_0(980)$ in the variable $s_{12}$ and $\sigma$ in $s_{13}$. The left and right
hashed regions correspond to   $\mid {\cal A}( m_0^2 - \epsilon, s_{13} ) \mid^2$
  and to $\mid {\cal A}( m_0^2 + \epsilon, s_{13}) \mid^2$ respectively}
\label{dalitz}
\end{figure}

The method proposed here  regards  the Dalitz plot just in the region
where a well established resonance, represented by a Breit-Wigner, for example, 
in $s_{12}$  spectrum
  \footnote{ In the three body decay $D\to i j k$, $s_{ij}$ and $s_{jk}$ 
  are two of the three possible Dalitz plot variables, that is 
  the square of the invariant mass of the particles $i$ plus $j$ and $j$ plus $k$
  respectively.},  crosses the complex amplitude under study in the $s_{13}$
  variable.
  In Figure \ref{dalitz} we show a Monte Carlo simulation of a Dalitz plot
  of  two scalar 
 resonances, one the $f_0(980)$ in the variable $s_{12}$ and the other 
 a broad and  low mass  resonance compatible with the $\sigma$ in $s_{13}$. 
 We show  that the interference created  in this particular region of the 
 phase space, can reveal the unknown  phase  variation of the amplitude
  as a function of $s_{13}$. The method can be applied to retrieve
  slow  variation similar to the one observed in 
  $K \pi$   elastic scattering at low invariant mass \cite{lass}, as well as 
  rapidly moving ones expected for resonant amplitudes.

  For simplicity we suppose that the only contributions
  we have in the Dalitz plot are the amplitude we want to study in  
  $s_{13}$ and a  scalar state represented by a constant width 
  Breit-Wigner in $s_{12}$.  This is not a limitation of the method, it 
  being straightforward to include known effects like extra states 
  populating the region as well as angular momentum distributions 
  and varying width Breit-Wigner functions. In this simple case, the 
 total amplitude is given  by:

\begin{equation}
{\cal A}(s_{12},s_{13}) = a_R {\cal BW}(s_{12}) 
+ a_s sin \delta (s_{13}) e^{i(\delta(s_{13})+\gamma)}
\end{equation}

\noindent
$\gamma$ represents the constant FSI phase difference  between the 
two body scattering amplitude and the bachelor particle.  $a_R$ 
contains  all constant or almost constant parameters, 
like form factors and the magnitude.  
$a_s e^{i\delta(s_{13})} sin \delta(s_{13})$ is 
the most general form to represent a complex amplitude, $a_s$ can be
considered constant in the range of the invariant mass region of interest. 
 The  relativistic Breit-Wigner function used in many
Dalitz plot analyses is given by:

\begin{equation}
{\cal BW}(s_{12}) = {m_0 \Gamma_0 \over {s_{12} - m^2_0 + im_0\Gamma_0}} 
\end{equation}
 
\noindent
$m_0$ $\Gamma_0$   are mass and the width of the resonance. 
The square modulus of the above amplitude can be written as:

\begin{eqnarray} 
\mid{\cal A}(s_{12},s_{13})\mid^2  = a_R^2 \mid {\cal BW}(s_{12})\mid^2  
+ a_s^2 sin^2\delta(s_{13}) +  \nonumber \\  \nonumber \\ {{2 a_R a_s m_0 \Gamma_0 sin\delta (s_{13}) 
\over  (s_{12}-m_0^2)^2 + m^2_0 \Gamma_0^2} 
((s_{12}-m_0^2) cos(\delta(s_{13}) + \gamma) + m_0 \Gamma_0 sin(\delta(s_{13}) + \gamma)}
\end{eqnarray}

In this case, $\mid {\cal BW}(s_{12})\mid^2$ is a   symmetric
function and thus 
${\cal BW}(s_{12} = m_0^2 + \epsilon )$ equals 
  ${\cal BW}(s_{12}' = m_0^2 - \epsilon )$, where $\epsilon$ is a small $s_{12}$
  interval. Consequently the difference of the amplitudes square takes the simple 
  form:
\begin{eqnarray}
\mid {\cal A}( m_0^2 + \epsilon, s_{13} ) \mid^2 
-\mid {\cal A}( m_0^2 - \epsilon, s_{13}) \mid^2  = 
{{4 a_s a_R \epsilon m_0 \Gamma_0 \over\epsilon^2 + m_0^2\Gamma_0^2} sin \delta(s_{13}) cos(\delta(s_{13})+ \gamma).} 
\end{eqnarray}

\noindent
 $\mid {\cal A}( m_0^2 - \epsilon, s_{13} ) \mid^2$ and 
$\mid {\cal A}( m_0^2 + \epsilon, s_{13}) \mid^2$ are taken from data as becomes
clear from  Figure
\ref{dalitz}.  We can rewrite Eq. 4 as:
\begin{eqnarray}
\Delta \mid{\cal A}\mid^2 =
\mid {\cal A}( m_0^2 + \epsilon, s_{13} ) \mid^2 
-\mid {\cal A}( m_0^2 - \epsilon, s_{13}) \mid^2  = 
 {\cal C} ( sin(2 \delta(s_{13})+ \gamma) - sin \gamma).
 \end{eqnarray} 
\noindent
 ${\cal C}$ is equal to $4 a_s a_R \epsilon m_0 \Gamma_0 /(\epsilon^2+m_0^2\Gamma^2_0)$. 
  $\gamma$ being a constant,
 $\Delta \mid{\cal A}\mid^2$ directly reflects the behavior of $\delta(s_{13})$.
   A constant 
 $\Delta\mid{ \cal A}\mid^2$ would imply constant $\delta(s_{13})$, this would
 be the case of  non-resonant contribution. The same way a slow phase motion
 will produce a slowly  varying $\Delta\mid{\cal A}\mid^2$ and a full resonance
 phase motion produces a clear signature in $\Delta\mid{\cal A}\mid^2$ with the
 presence of zero, maxim and minim values.
 Next we discuss how to extract the constant phase $\gamma$
 and $\delta(s_{13})$ for two situations.  We start with the case of a rapidly moving phase
 like a resonance where we extract direct information on $\delta(s_{13})$. Then
 we deal with  slow  phase variation, for which  we use as the example the
 LASS $K \pi$ low mass phase behavior.

To study a resonant  amplitude  in the  $\Delta\mid{\cal A}\mid^2$
distribution, we produced  a  fast simulation of 10K events with a hypothetical 
decay \d3pi with only a broad and low mass scalar resonance $\sigma$ in $ s_{13} $  and 
the well known resonance   $f_0(980)$ in $ s_{12} $ and strong phase difference
between the two of 2.8 rad, Figure \ref{dalitz} shows  the relevant portion of
the  Dalitz plot.
To generate the  $f_0(980)$ we used the constant width Breit-Wigner (Eq. 2).
For the low mass wide $\sigma$, the Breit-Wigner parameterization produces
undesirable threshold features. We could, in
principle, introduce form-factors to avoid this problem, but this would bring 
unnecessary complexity to the example. Instead ,we 
 parameterized the low mass  resonance as 
 $ sin\delta(s_{13}) e^{i\delta(s_{13})}$, like the amplitude we used in Eq. 1. 
 with $\delta(s_{13}) = \pi$ / (1+ exp (a($s_{13}-m_0^2$)), the parameter a is 
 associated with the $\sigma$ width. We used $ a = -15$ GeV$^{-2}$. With this choice, we 
 have a complete  $\pi$ phase variation around the $\sigma$ mass, as would 
 behave a Breit-Wigner far from the threshold, and the mass plot given by
 $ \mid{sin\delta(s_{13}) e^{i\delta(s_{13})}}\mid^2$ reproduces the overall
 characteristics of the $\sigma$ resonance.

\begin{figure}[hbt]
\epsfxsize=43pc
\centerline{
\epsfbox{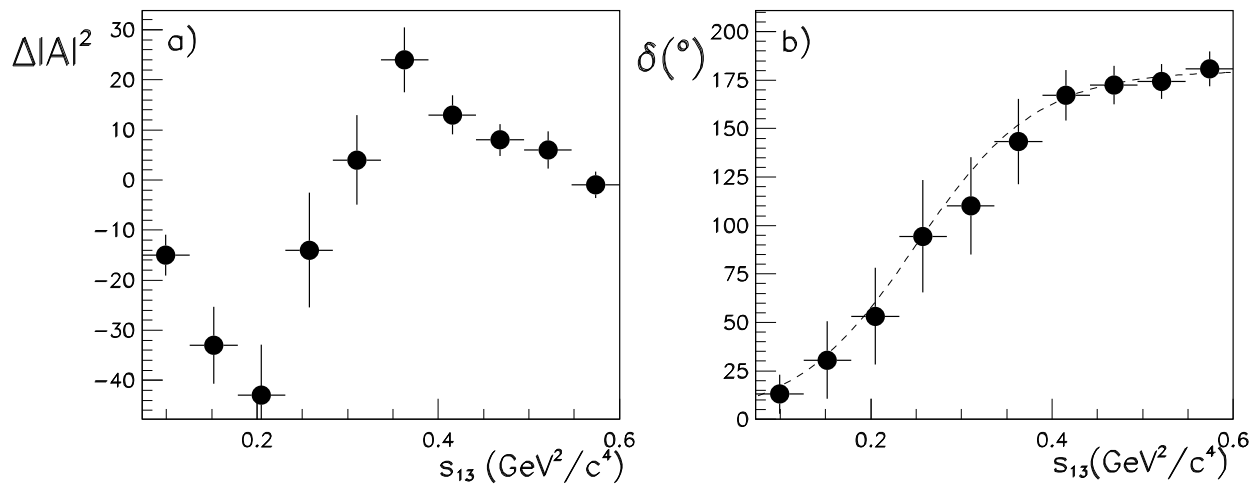}}
\caption{a) $\Delta\mid{\cal A}\mid^2$ distributions for resonance 
$\sigma$ like interfering with $f_0(980)$ generated with a 10K fast-MC events.
 b)$\delta(s_{13})$ distribution 
got from $\Delta\mid{\cal A}\mid^2$ distribution presented in a). The dashed
line is the function used as input to the fast-MC. }
\label{sigma}
\end{figure}

 The $\Delta\mid{\cal A}\mid^2$ distribution for the fast-MC produced with the 
above parameters is presented in Figure \ref{sigma}a. We can see a strong 
variation of this  distribution with the presence of three 
zeros:  one at the threshold implicit from Eq. 5, the second at 
$s_{13}\approx 0.3$ associated with
$\delta = 3\pi/2 - \gamma$ and the third at $s_{13}\approx 0.55 $
 corresponding to 
 $\delta = \pi $. Thus, it is possible to observe in a simple way, 
 a clear signature of the presence of large phase 
  variation associated with  the number of zeros in $\Delta\mid{\cal A}\mid^2$.  
 
\begin{figure}[hbt]
\epsfxsize=22pc
\centerline{
\epsfbox{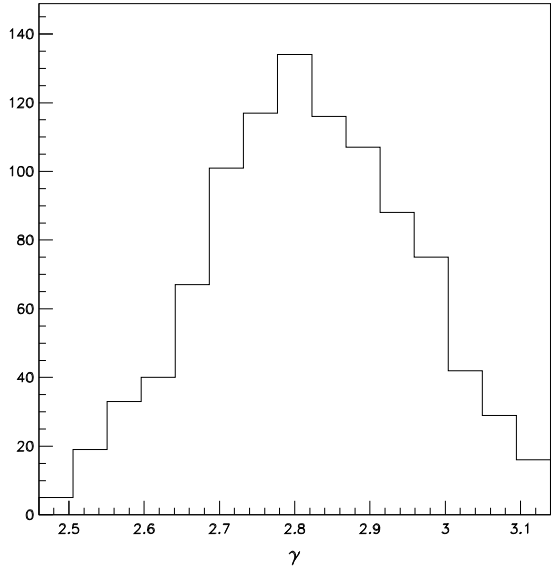}}
\caption{$\gamma$  distribution for  1000 fast-MC experiments. 
phase 2.8 rad. }
\label{rms}
\end{figure}

 We can extract directly the phase motion 
 $\delta(s_{13})$ noticing that the maximum and minimum values of  
 $\Delta\mid{\cal A}\mid^2$ corresponds respectively to the 
  $sin(2\delta(s_{13})+ \gamma) = 1$ and $sin(2\delta(s_{13})+ \gamma) = - 1$
  conditions. These and the $\Delta\mid{\cal A}\mid^2$ value, contents 
  of bins 3 and 6 of the plot Figure \ref{sigma}a provide us two
  equations that can be used to get $\gamma$ and  ${\cal C}$.
   Then we invert  Equation 5 and 
get from each value of  $\Delta\mid{\cal A}\mid^2$ one value of 
$\delta(s_{13})$. Figure \ref{sigma}b compares $\delta(s_{13})$ obtained in 
this way with the functional form (dashed) used to generate the fast-MC sample.   
To evaluate error in $\delta(s_{13})$ (Figure \ref{sigma}b), and the  
ability of this procedure to measure the phase $\gamma$ we repeated the 
fast-MC experiment 1000 times. The result is shown in Figure ~\ref{rms}.  
We are able to recover the input value (2.8 rad) and the rms of the 
distribution is consistent with the individual experiment errors.

We have presented an ideal situation in our example, with constant width, 
and isolated scalar resonances. In a realistic case, the width dependence
introduces an asymmetry in $s_{13}$, the same  occurs with  respect to 
resonant amplitudes with spin 1 or 2 and even form factors \cite{bugg}. 
These effects can be included 
in Equation 3 in a complete way, since we know very well the functional 
form of a Breit-Wigner, the width dependence with $s_{12}$, and the spin 
amplitude. However, the  simplified amplitude 
for $\Delta\mid{\cal A}\mid^2$ still works if we are willing to sacrifice the
measurement of  $\gamma$. In fact, we observed that many 
of these {\it realistic} effects are absorbed by the parameter  $\gamma$,
 keeping the $\delta(s_{13})$ distribution almost unchanged if we include  
 or not the  {\it realistic}  effects. Since with this method, one is primarily
interested  on the direct measurement of the phase motion, and  since  
 the usual Dalitz plot method has been the best  technique to 
 get the FSI phase, one can use  $\Delta\mid{\cal A}\mid^2$ 
without much elaboration in most cases.  

\begin{figure}[hbt]
\epsfxsize=38pc
\centerline{
\epsfbox{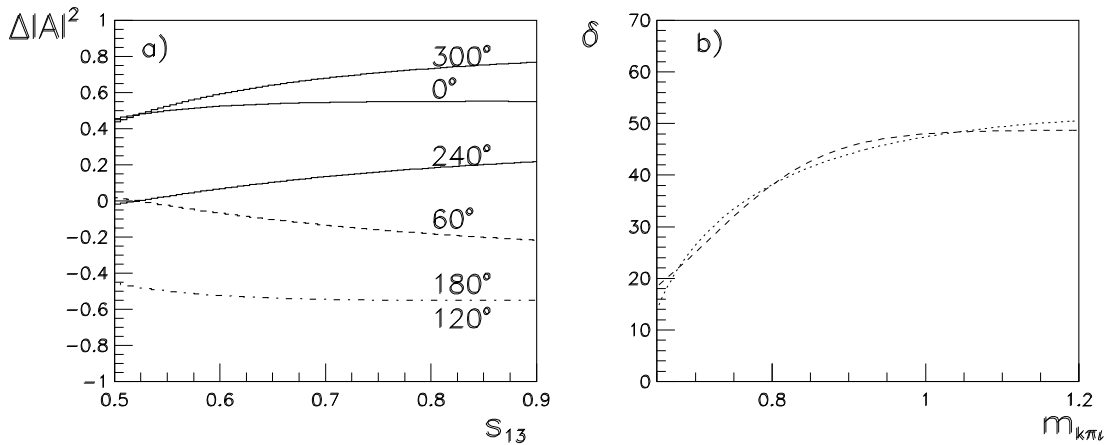}}
\caption{(a)$\Delta\mid{\cal A}\mid^2$ distributions for different 
$\gamma$ values. (b)  LASS low mass phase motion used as input (dotted line)
and obtained fitting 
 $\Delta\mid{\cal A}\mid^2$ distribution using the functional form of
 $\delta(s_{13})$ given in text (dashed line).  }
\label{lass}
\end{figure}

The other case that we wish to discuss is when the 
 complex amplitude is not a resonance, but has only a  
small phase variation. A good example is   the  scalar smooth phase 
variation at low mass obtained 
by the LASS \cite{lass} collaboration in $ K \pi$  scattering. The LASS phase variation is given by 
$ cot \delta = 1/d.q + e.q /2$  \cite{lass}, where $q$ is the momentum in 
the $K^- \pi^+$ center  of mass, $d=4.03\pm 1.72 $ GeV$^{-1}$ and  
$e=1.29 \pm 0.89$ GeV$^{-1}$ are fit parameters.  We  included this amplitude 
as a function of $s_{13}$ in a hypothetical three body charm decay \kpipi 
with the well known resonance $K_0(1430)$  in the crossing channel, 
$ s_{12}$. Figure   \ref{lass}a shows the $\Delta\mid{\cal A}\mid^2$  
behavior, given by Eq. 5,  for several values of $\gamma$ between $0^0$ 
to $300^0$.  While all curves show slow  variation of 
$\Delta\mid{\cal A}\mid^2$, we notice the clear distinction
 among the  $\gamma$ values. 
 
 In this case we can not extract directly the phase motion 
 since we have not maxima and minima, as we have  for the resonance example.
 Instead we have to use a functional form to fit data and get the 
 shape. The fit function should be  monotonic with a constant 
asymptotic behavior.  We used for the  $\delta(s_{13})$ form the 
 same function we used to generated  the resonance in the preceding example, 
 that is $\delta(s_{13}) = ph $ / (1+ exp (a($s_{13}- t$)),
 where $ph$, $a$ and $t$ are free fit parameters. With this simple function we
  are able to retrieve the $\delta(s_{13})$ behavior and $\gamma$ values.
 In Figure \ref{lass}b we show  the LASS phase behavior 
\cite{lass}, input to the exercise and the dashed line  the functional 
form we get with our fit function.

We have presented here a method to observe  phase motion of a complex 
amplitude in a Dalitz plot of three body  decays. We showed that the    
difference of amplitudes, $\Delta\mid{\cal A}\mid^2$, around the
 central mass squared of a well known resonance  is a good way to 
 observe this variation.
We also showed that one can determine   rapid or even slow phase
variation with this method. We used as example two scalar resonances,
  the $\sigma$ and the $f_0(980)$ in 
\d3pi decay. This is the  simplest possible situation,  
however one could use also the crossing between the $\sigma$ and the 
$f_2(1270)$ resonance including the well known angular distribution. 
The advantage of using $f_2(1270)$ in real data is the large statistics
in the \d3pi decay, associated with  the size of 
the branching fraction and the effective area of the crossing region. 
Nevertheless,  we believe that both $f_0(980)$ and $f_2(1270)$ could be used to measure 
the phase motion of the low $\pi^+\pi^-$ invariant mass region. The analysis of the 
$\kappa$ resonance in \kpipi decay, should be slightly more complicated 
because of the proximity of the vector resonance $K^*(890)$ to the 
new scalar state. However the $K^*(890)$ would have to be 
incorporated in Eq. 1.

We thank  Professors Jeff Appel, Hans G\"uenter Dosch,  Wolfgang Ochs and 
Alberto Reis for suggestions and important comments.

\end{document}